\documentclass[aps,preprint]{revtex4}
\usepackage{graphicx}

\begin{document}
\title{What governs the fluidic behavior of water near single DNA molecules
at the micro/nano scale \footnote{Supported by  the National
Natural Science Foundation of China under Grant No. 10474109,
Foundation of Ministry of Personnel of China and Shanghai
Supercomputer Center of China} }

\author{Yi Zhang$^{1}$, Huabing Li$^{1}$, Xiaoling Lei$^1$,\\
        Junhong Lv$^1$, Xiaobai Ai$^1$,\\
         Hu Jun$^{1,2}$\footnote{Corresponding author. Email:
jhu@sjtu.edu.cn, syc@me.jhu.edu, fanghaiping@sinap.ac.cn}, Shiyi
Chen$^{3,\dagger}$ and Fang Haiping$^{1,\dagger}$}
\affiliation{$^1$Shanghai Institute of Applied Physics, Chinese
Academy of Sciences, P.O. Box 800-204, Shanghai 201800, China\\
          $^2$Bio-X Life Sciences Research Center,
           Shanghai JiaoTong University, Shanghai 200030, China\\
          $^3$Department of Mechanical Engineering, the Johns Hopkins University,
          MD 21218, USA
          }


\begin{abstract}

The fluidic behavior of water at the micro/nano scale is studied
by using of single DNA molecules as a model system. Stable curved
DNA patterns with spans about one micron were generated by using
of water flows, and observed by Atomic Force Microscopy. By
rigorously comparing the numerical simulation results with these
patterns, it is suggested that the form of the macroscopic
hydrodynamic equation still works quantitatively well on the fluid
flows at the nanoscale.  The molecular effects, however, are still
apparent that the effective viscosity of the adjacent water is
considerably larger than its bulk value. Our observation is also
helpful to understand of the dynamics of biomolecules in solutions
from nanoscale to microscale.
\end{abstract}
 \maketitle

 Novel micro-fluidics has provided new possibilities for the
development of fabricating high-performance devices with
micrometer or even submicrometer dimensions
\cite{Volkmuth92,Ho01,Hu02}, as well as bio-processing for
bio-medical analysis \cite{Thorstenson98}. The ability to control
accurately the flow requires the understanding of fluid flows at
the micro/nano scale near macromolecules or above solid surfaces.
Recent experiments have shown that macroscopic hydrodynamics can
reliably describe the flow of fluids through channels with
cross-sectional dimensions that range from tens to hundreds of
micrometers \cite{Truskett03}. The nanometer scale, however, is in
the transition regime between continuum and molecule dominated
conditions \cite{Ho01,Laughlin00}, about which we know very little
\cite{Laughlin00}.  Molecular dynamics has been used to study the
nanofluid \cite{Koplik88} and consistence between the numerical
results from molecular dynamics and Navier-Stokes equations has
been obtained recently \cite{Nie03}. However, both high and normal
`effective' viscosities, its departure from the bulk value usually
characterizing the molecular effects, and boundary slip have been
observed in the confined water when the sample thickness is
comparable to or below the diameter of DNA molecules \cite{Zhu01},
{\it indicating that the dynamics of nanofluid should be much
complex}. Moreover, it has been found experimentally that the
water flow through orifices with diameter about 8 $\mu$m is
considerably different from that of ordinary size
\cite{Hasegawa97}. At a hydrophobic surface, in an unexpectedly
extended ($\sim$ 4 $nm$) water layer, the density of water has a
noticeable reduction (85-90$\%$ of the density of bulk water)
\cite{Schwendel03}. Consequently, the fluidity of water several to
tens of nanometers above non-liquid materials should be different
from that of the macroscopic systems despite the fact that the
water molecules are much smaller.

 The difficulty in the experimental studies is partially
due to the limited experimental technologies to obtain data that
can be compared by theoretical predictions quantitatively.
Although molecules with fluorescent labels have been widely used
\cite{Smith99} to make the biomolecules visible, the resolution is
limited by the minimal photon wavelength, which is about half a
micron.
 Atomic Force Microscopy (AFM) and Electronic Microscopy
(EM) have the nanometer resolution.  However, they require the
objects to be fixed on solid substrates. It is obviously
questionable that the quantitative details on the dynamics can be
exploited from motionless objects unless they are in
steady-states, which are difficult to be confirmed.

The study of biological systems will continue to inspire the
development of new physics \cite{Frauenfelder99} and the dynamics
of DNA in fluid flow has been extensively studied by various
methods \cite{Smith99,Meiners00}. Thanks to the elastic behavior
of DNA fibers that they reach maximal lengthes with a tension
larger than $\sim$65 pN \cite{Cluzel96} and are not broken until
400 pN \cite{Bensimon95}, those overstretched DNA fibers can reach
steady states in the fluid flows and be fixed on substrates. In
this Letter a simple experiment was then set up to obtain steady
curved DNA patterns for span lengths about or less than 1 $\mu$m,
generated by water flow. By rigorously comparing the observed DNA
patterns with theoretical predictions, it is suggested that
macroscopic hydrodynamic equations still works quantitatively well
on the fluid flow at the nanoscale at least at steady states. To
our best knowledge, this is the first report on the applicability
of the Navier-Stokes equations on the fluidic behavior of water
near single DNA molecules at the nanoscale by testing the
numerical predictions with the experimental observation.

The $\lambda$DNA, purchased from Sigma Co. (USA), was diluted
before use to a concentration of 3 ng/$\mu$l with TE buffer (40
mMol/L Tris-HCl, 1 mMol/L EDTA, pH 8.0). The experiment began by
depositing a drop of 10 $\mu$l $\lambda$DNA solution onto a mica
sheet modified by APTES \cite{Lu04} that was placed on a rotary
device. The distance between the drops of $\lambda$DNA solution
and the axis of the rotary device was always about 10 mm. The spin
speed of rotation is kept to be 1150 rpm. After rotation, the drop
would be thrown off and some DNA molecules were aligned on the
surface. To demonstrate that DNA molecules were aligned by water
flow rather than the liquid-gas interface, some DNA fibers were
observed by Tapping-Mode AFM (Nanoscope IIIa, Veeco Instruments,
Santa Barbara, CA) using silicon tips (Silicon-MDT Ltd., Russia)
\cite{Hu02}. It is found that the relative extensions of those
aligned DNA molecules are within $100\% \pm 5\%$ while those are
about $133\%$ to $150\%$ in the molecular combing by the receding
meniscus \cite{Bensimon94,Bensimon95}. Then another drop of water
was placed on the same area of the surface on which there are
aligned DNA molecules while the orientation of the mica sheet on
the rotary device was changed about 90$^{\rm o}$. A subsequent
rotation created a second flow on the mica surface, approximately
perpendicular to the first one. The velocity of the drop was
estimated to be about 1 cm/s, with an upper limit of 3 cm/s. The
as-generated DNA patterns were revealed by Tapping-mode AFM
\cite{Hu02}.

\begin{figure}[tbh]
    \begin{center}
         \scalebox{1.2}[1.2]{\includegraphics*[10,20][320,230]{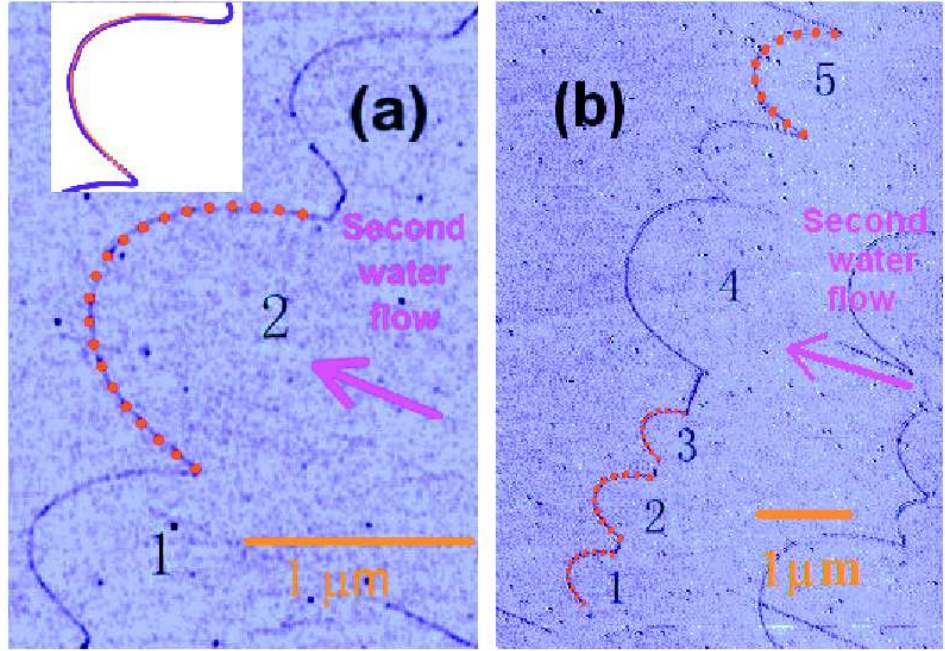}}
    \end{center}
\caption{ DNA patterns obtained by AFM imaging. Some of the DNA
segments still keep their original aligned patterns, while some
segments bend to the moving direction of liquid flow and form
wave-like patterns. The curves denoted by red dots are the
theoretical predictions $\Re$ from our numerical calculation which
agree extremely well with the curves from the experiment.  The
light magenta arrows are only served as a guide to eyes rather
than the exact directions of the second water flows. The blue dots
in the inset are the data extracted from the DNA pattern and the
red line is the theoretical prediction $\Re$.} \label{bound}
\end{figure}
We have systematically observed over 100 curves with span length
$L$, 50 nm $<L<$ 2 $\mu$m after the second water flow, as
typically shown in Fig. 1. Some of the DNA segments still have
their original aligned patterns, while other segments bend in the
direction of the second water flow. Among them, 7 symmetric curves
are constituted of those overstretched DNA fibers, which reach
their maximal lengthes, with span length ranging from $\sim$580 nm
to $\sim$1260 nm. The fact that the ratio of the contour and the
span is about 1.75 for those overstretched patterns further
verifies that DNA molecules are aligned and curved by water flows
rather than the interface. The relative extension is calculated
from the contour and the span length $L$ of a curve. In the
following we will demonstrate that these symmetric and steady DNA
curves are a result of the balance of the macroscopic hydrodynamic
forces with an adjusted viscosity, as illustrated by the 5 typical
examples shown in Fig. 1, i.e., curve 2 in (a) and curves 1, 2, 3,
5 in (b). The other patterns, including the asymmetric curve 1 in
Fig. 1(a) and the double-peaked curve 4 in Fig. 1(b), have been
initially analyzed in \cite{Li03} and will be discussed in details
elsewhere.

The numerical analysis is based on the bead-spring model
\cite{Larson99}. To test whether the macroscopic hydrodynamic
equations still works quantitatively well, the forces acting on
the beads are computed by simulating water flow according to the
Navier-Stokes equations. Explicitly, the lattice Boltzmann
simulation \cite{Chen91} is applied to simulate the water flow
around the spherical beads. A non-slip boundary condition
\cite{Filippova97} is applied at the bead-fluid boundaries. The
momentum exchange \cite{Ladd94,Li04} at the solid-fluid interface
is used to determine the force acting on the DNA beads. In the
lattice Boltzmann simulations, the non-dimensional radius of each
bead is 5.2 lattice units and the relaxation time $\tau=0.75$. The
density and kinetic viscosity of the water are 998.3$kg/m^{3}$ and
1.007$\times 10^{-6} m^2/s$ \cite{Douglas}, respectively.

\begin{figure}[tbh]
    \begin{center}
        \scalebox{1.4}[1.4]{\includegraphics*[15,30][210,195]{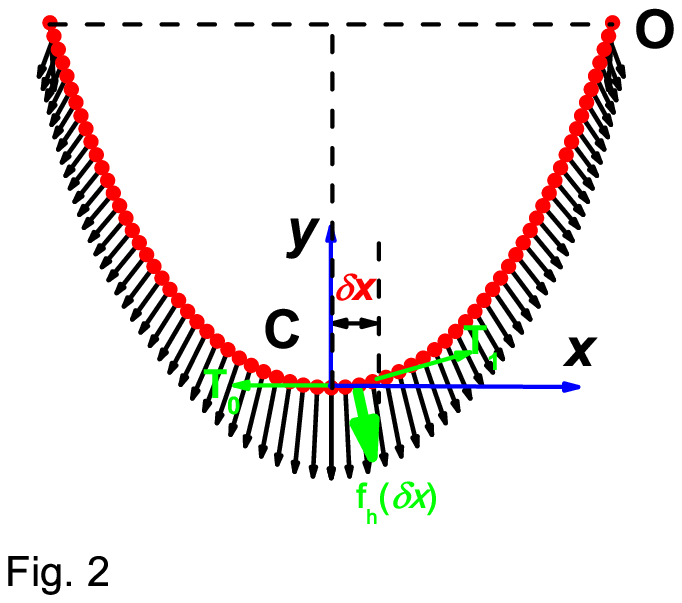}}
    \end{center}
    \vspace{-0.8cm}
\caption{The system for our numerical simulation together with the
coordinates. The red circles are the beads placed on the DNA
pattern scaled down from the curve marked by 2 in Fig. 1(a). The
black arrows represent the directions and magnitudes of the
hydrodynamic forces on the beads obtained by numerical
simulations. There are three forces acting on the DNA segment
between $C$ ($x = 0$) and $x = \delta x$: the tension {\bf T}$_C$
at $C$ along the $x$-axis, the tension {\bf T}$_1$ at the other
end of the segment and the hydrodynamic force {\bf f}$_h (\delta
x)$ on this segment, which can be computed from the hydrodynamic
force distribution.} \label{bound}
\end{figure}

Let us take the curve marked by 2 in Fig. 1(a) as an example. The
settings of our numerical simulations are shown in Fig. 2. Some
beads, say sixty-nine, were placed on the DNA pattern scaled down
from the curve.  The radius of each bead is 1 nm, approximately
equal to that of the DNA strands, while the center-to-center
distance between the nearest-neighbor beads is always 2.308 nm.
Periodic boundary conditions are applied for the $x$ and $z$
directions. The mesh sizes are: $L_x$ is a little larger than the
span length $L$ of the DNA pattern, the inlet and outlet in $y$
direction is always 30.5 nm from the closest bead, see, e.g.,
$L_x=95.5$ nm and $L_y=120.0$ nm for $L$=90.9 nm; $L_z$=41.1 nm.
It is found that $L_z$ is large enough so that the effect of the
periodic boundary condition in the $z$ direction is negligible.
The fluid velocity at the inlet and outlet is assumed to be 1 cm/s
and along the -$y$ direction at all time.

 The force distribution acting on beads is shown in Fig. 2.
Since the actual DNA fiber is continuous rather than discrete, we
therefore used the best continuum curve fitting from the
calculated discrete force to approximate the realistic {\it
hydrodynamic force distribution} along the DNA fiber.

It is evident that if a DNA fiber is in a steady state of the
hydrodynamic force, the total force, i.e., the sum of the
hydrodynamic force and the tensions, acting on any segment of the
DNA fiber should vanish. Unfortunately, the tension is unknown and
it is difficult to measure the velocity of water flow acting on
the DNA molecules to an acceptable accuracy so that we cannot
examine the equilibrium condition explicitly. However, if we
assume that the coordinates of the two ends of the DNA at
locations $C$ and $O$ (see Fig. 2) are given, then the {\it
hydrodynamic force distribution} obtained above will uniquely
determine a curve $\Re$. Further, $\Re$ can be used to compare
with the experimental result. If these two curves coincide with
each other, the DNA pattern obtained by the experiment is indeed
in the steady state of the hydrodynamic force. Otherwise the
discrepancy reflects the special behavior at the nanoscale.

To construct the curve $\Re$, we start our steady force balance
calculation from the critical point $C$ as shown in Figure 2.
Assuming that there is a small displacement $\delta x$ in the
$x$-direction, then there are three forces acting on the DNA
segment between $C$ ($x = 0$) and $x = \delta x$: the tension {\bf
T}$_C$ at $C$ along the $x$-direction, the tension {\bf T}$_1$ at
the other end of the segment and the hydrodynamic force {\bf f}$_h
(\delta x)$ on this segment. $\delta x=0.5$ nm was used in our
calculations. For the equilibrium state, the force balance for the
DNA segment gives:
\begin{equation}
{\bf T}_C+{\bf T}_1+{\bf f}_h(\delta x)=0. \label{balance}
\end{equation}
If we assume that the magnitude of $|{\bf T}_C|$ is $T_C$, both
the magnitude and direction of the tension {\bf T}$_1$ can be
obtained through Eq. \ref{balance}. Consequently we can obtain the
$y$ coordinate of the DNA segment at $x=\delta x$. Next, we
consider the position of the curve $\Re$ at $x=2\delta x$. The
tangential direction at this point can be obtained similarly to
find the increment of $y$ value from $x=\delta x$ to $x=2\delta
x$. This extrapolation process can be repeated until $x_o$ is
reached. Then an iterative process is used to self-consistently
determine $T_C$ to meet the value $y_o$ required at $O$. In this
way, we obtain the curve $\Re$. It should be noted that the force
to bend the DNA molecules to the curves shown in Fig. 1, except
for the parts near the anchored points, is about 0.1 pN,
negligibly smaller than the tension in the DNA curves.

$\Re$ is displayed as red circles in Fig. 1(a). To our great
surprise it is found that $\Re$ from our theoretical prediction
agrees extremely well with the curve from the experiment.
Numerical simulations for $35, 49$ and $89$ beads, corresponding
to span lengths $L=45.1$, 63.6, and 116.6 nm, respectively, have
also been carried out. It is found that the normalized patterns
from numerical predictions with different length scales can be
normalized to a high accuracy. In the same way, we obtain the
theoretical predictions $\Re$ for the curves 1, 2, 3, 5 in Fig.
1(b). In order to obtain a quantitative comparison between the
numerical predictions and the experimental results, the global
relative error is computed from those 5 curves, which is 5 $\pm$ 1
$\%$.

\begin{figure}[tbh]
    \begin{center}
        \scalebox{1.6}[1.6]{\includegraphics*[10,18][180,165]{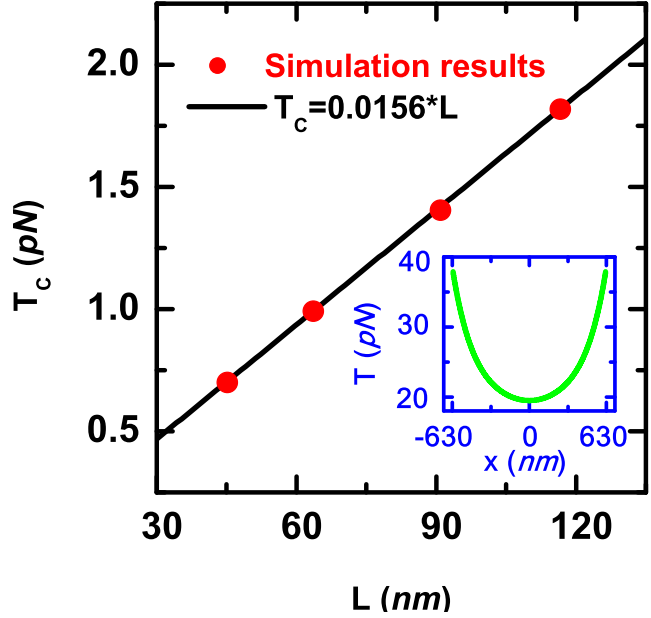}}
    \end{center}
\caption{ The tension $T_C$ at the point $C$ of the DNA curve
shown in Figure 2 for different span lengths, and their linear fit
function. The inset is the tension distribution along the DNA
fiber for the span length of 1260 nm, computed from a bulk
viscosity. The tension at $C$ has its minimum value and the
tension at $O$ has its maximum value $T_O \approx 1.9 T_C$.}
\label{tension}
\end{figure}

The tension $T_C$ at the critical point $C$ from our calculations
for different $L$ is shown in Fig. \ref{tension}, which fits very
well to a linear function $T_C=0.0156L$ when the inlet fluid
velocity equals 1 cm/s. Following this linear scaling, we found
that $T_C=$ 9.0 and 19.7 pN for $L=580$ and 1260 nm, respectively.

The tensions are larger than 65 pN \cite{Cluzel96} in the
overstretched DNA fibers with relative extension about 1.75.
However, $T_C$ calculated for $L=580$ nm is only 9.0 pN. Even for
the upper limit of the velocity, 3 cm/s, of the water drop,
$T_C=27.0$ pN is still smaller than 65 pN. We also note that the
viscosity of water is the same as its bulk value in the
calculation. If we assume that the local effective viscosity near
the DNA fiber is more than 3 times of its bulk value, the
calculated tension in all the 7 steady and symmetric curved DNA
patterns will be larger than $65$ pN. Enhanced by a factor of 6
for the effective viscosity of water near proteins has been also
observed recently \cite{Halle03}. In the inset of Fig.
\ref{tension}, we have included the tension distribution along the
DNA fiber for $L$ of 1260 nm.

The numerical analysis in the manuscript has neglected the fact
that DNA fibers are close to the mica surface. It has shown that
polymer molecules can move on the surface with a low barrier
although the absorption on molecules by the surface is strong
\cite{Gennes87}. Therefore, we believe the interaction between the
extended DNA strands and the substrate will not change the
conclusion we obtain above. In this case the flow around the DNA
fibers may be a shear flow. We have performed new numerical
simulations. The distance from the $x-y$ plane, where the DNA
curve located, to a non-slip boundary is $\zeta$. The velocity of
the flow at the inlet increases linearly from the boundary whereas
the velocity in the line with the $x-y$ plane is 1 cm/s. The other
settings are the same as before. It is found that the normalized
patterns $\Re$ obtained in this way for $\zeta=3.84$ nm and $0.58$
nm are consistent with those obtained above perfectly, while the
tensions increase by a factor of 1.2 and 2.6 separately.  The
fluid velocity near wall is usually much smaller than that of the
drop of water, therefore higher effective viscosity is also
demanded.

The accuracy of the fitting of the theoretical predictions to the
experimental patterns is at the level of AFM resolution, about
10-20 nm. This observation suggests that the force exerted by the
adjacent water on the DNA fiber can be determined quantitatively
by the macroscopic hydrodynamic equations at least in the steady
state even at a resolution of about 20 nm. The viscosity of the
adjacent water, however, is much larger than its bulk value. It
should be noted that the excellent agreement between our numerical
predictions and the experimental results further demonstrate the
robustness of the numerical model although the detail of the
experiment should be much complex.

We thank Dr. Gary D. Doolen, Profs. D. Bensimon, V. Croquette and
Z. Ye for helpful discussion and suggestions, Miss Chunmei Wang
for extracting DNA data from AFM images.

\end{document}